# Objective comparison of auditory profiles using manifold learning and intrinsic measures

Chen Xu[a]*, Birger Kollmeier[a], and Lena Schell-Majoor[a]

[a]*Medizinische Physik and Cluster of Excellence Hearing4all, Universität Oldenburg, D-26111 Oldenburg, Germany*

Contact: Chen Xu

chen.xu@uni-oldenburg.de

Department of Medical Physics and Acoustics, Faculty VI

University of Oldenburg, 26111, Oldenburg, Germany

Objective comparison of auditory profiles using manifold learning and intrinsic measures

Assigning individuals with hearing impairment to auditory profiles can support a better understanding of the causes and consequences of hearing loss and facilitate profile-based hearing-aid fitting. However, the factors influencing auditory profile generation remain insufficiently understood, and existing profiling frameworks have rarely been compared systematically. This study therefore investigated the impact of two key factors—the clustering method and the number of profiles—on auditory profile generation. In addition, eight established auditory profiling frameworks were systematically reviewed and compared using intrinsic statistical measures and manifold learning techniques. Frameworks were evaluated with respect to internal consistency (i.e., grouping similar individuals) and cluster separation (i.e., clear differentiation between groups). To ensure comparability, all analyses were conducted on a common open-access dataset, the extended Oldenburg Hearing Health Record (OHHR), comprising 1,127 participants (mean age = 67.2 years, SD = 12.0). Results showed that both the clustering method and the chosen number of profiles substantially influenced the resulting auditory profiles. Among purely audiogram-based approaches, the Bisgaard auditory profiles demonstrated the strongest clustering performance, whereas audiometric phenotypes performed worst. Among frameworks incorporating supra-threshold information in addition to the audiogram, the Hearing4All auditory profiles were advantageous, combining a near-optimal number of profile classes (N = 13) with high clustering quality, as indicated by a low Davies–Bouldin index.

In conclusion, manifold learning and intrinsic measures enable systematic comparison of auditory profiling frameworks and identify the Hearing4All auditory profile as a promising approach for future research.

**Introduction**

Traditional hearing loss classifications based on pure-tone averages (e.g., the World Health Organization (WHO) terminology of a "moderate" hearing loss Humes (2019)) often fail to capture the diversity of hearing impairments, necessitating completely different treatments for patients in the same class. To enable more personalized diagnostics and interventions, recent research has introduced finer-grained subgroups within the hearing-impaired population. These subgroups are defined using the concept of auditory profiles which will be bench-marked in the present paper using machine learning methods.

Auditory profiling is a key concept in Precision Audiology (Dreschler et al., 2008; Van Esch et al., 2013; Vlaming et al., 2011; Sanchez Lopez et al., 2018; Saak et al., 2022; Cherri et al., 2024), offering a promising solution for more accurate hearing loss diagnostics and treatment beyond the audiogram. On the one hand, it captures additional information, such as supra-threshold deficits (Kollmeier 1999; Sanchez Lopez et al., 2018), enabling more relevant diagnostics. On the other hand, it enables the development of profile-based hearing device fittings (see Sanchez-Lopez et al., 2021; Kubiak et al., 2022). Given these advantages, investigating auditory profiles may offer valuable insights on the interaction of different diagnostics tests and their relation to auditory functions most relevant for rehabilitative audiology.

Auditory profiling ultimately aims to support practical applications, such as improved diagnostics, individualized treatment, faster hearing-aid fitting, or targeted participant selection for research (Saak et al., 2022; 2025). Although these applications are not directly addressed in the present study, they motivate the need for a systematic, objective evaluation of auditory profiling frameworks. By focusing on intrinsic clustering properties and methodological design choices, this work provides a foundation for selecting profiling approaches that are appropriate for different application contexts.

The first type of auditory profiles are the classical "audiogram-based auditory profiles" solely generated from participants' audiograms. In its most simple form, participants are classified based on their pure-tone average across 0.5, 1, 2, and 4 kHz (PTA4). Individuals with a PTA4 below 20 dB HL are considered as normal hearing (NH), while those above this threshold are classified as hearing impaired (HI). This is used as the "baseline" auditory profile in the current study. The WHO provides a more detailed PTA4-based grading system—"normal," "mild," "moderate," "moderately severe," "severe," and "profound" (Humes, 2019).

More sophisticated audiogram-based auditory profiles consider the shape of the audiogram: Bisgaard et al. (2010) identified 10 classes of audiograms, also known as standard audiograms or Bisgaard profiles, which include seven flat or moderately sloping audiograms (N1 to N7) and three steeply sloping audiograms (S1 to S3) derived through vector quantization. Additionally, the Wisconsin Age-Related Hearing Impairment Classification Scale (WARHICS), introduced by Cruickshanks et al. (2020) and Humes et al. (2021), defines eight levels of hearing loss based primarily on audiometric thresholds, with each level corresponding to a distinct auditory profile. Dubno et al. (2013) proposed five auditory profiles, referred to as audiometric phenotypes, based on the audiogram of an animal model. These include older-normal (O-N), pre-metabolic (PRE-MET), metabolic (MET), sensory (SENS), and combined metabolic and sensory (MET + SENS) profiles. Building on this work, Parthasarathy et al. (2020) used data-driven methods to derive four general phenotypes: normal audiogram, flat sloping hearing loss, high-frequency hearing loss (HFHL), and mixed sensorineural hearing loss.

The second type of auditory profiles can be described as "comprehensive auditory profiles" since they are generated using not only audiograms but also supra-threshold parameters, such as speech recognition thresholds in noise, and employ sophisticated data-driven methods. The first comprehensive profile studies were performed in the "HearCom" project (Dreschler et al., 2008; Van Esch et al., 2013; Vlaming et al., 2011) which did not lead to a practically applicable set of auditory profiles, but was seminal to further work in this area: Sanchez-Lopez et al. (2018; 2020) developed four distinct auditory profiles as part of the Danish "BEAR" project, referred to as "BEAR profiles" in this paper, while Saak et al. (2022; 2025) identified 13 different profiles within the German 'Hearing4all' project, therefore referred to as "Hearing4all auditory profiles".

A summary of these auditory profiles, along with the ones reviewed earlier, is presented in Table 1. The table reports the number of profiles for each framework. It also highlights that only WHO hearing impairment (HI) grades and WARHICS levels are based on epidemiological datasets, whereas the other six auditory profiling frameworks were developed using more specific clinical auditory datasets. Additionally, the baseline, WHO HI grades, WARHICS levels, and audiometric phenotypes are derived using expert knowledge (also referred to as 'expert-based' approaches), while the other four auditory profiling frameworks rely on data-driven methods.

Table 1. Characteristics of the eight auditory profiling frameworks compared in this study. N = number of profiles. n = number of patients required for the generations of the auditory profiles.

| Auditory profiles | N | Parameters included | Development dataset | Profiling approach | Reference |
|---|---|---|---|---|---|

|  | Audiogram-based | Comprehensive | Epidemiological (n ≥ 4000 patients) | Audiological (n ≤ 1000 patients) | Expert-based | Data-driven |  |
| --- | --- | --- | --- | --- | --- | --- | --- |
| Baseline | 2 | × |  |  | × | × | - |
| Bisgaard profile | 10 | × |  |  | × |  | × | Bisgaard et al. (2010) |
| WHO HI grades | 6 | × |  | × |  | × |  | Humes (2019) |
| WARHICS levels | 8 | × |  | × |  | × |  | Cruickshanks et al. (2020) |
| phenotype | 5 | × |  |  | × | × |  | Dubno et al. (2013) |
| General phenotype | 4 | × |  |  | × |  | × | Parthasarathy et al. (2020) |
| BEAR | 4 |  | × |  | × |  | × | Sanchez-Lopez et al. (2020) |
| Hearing4all | 13 |  | × |  | × |  | × | Saak et al. (2022) |

Despite the advantages and significance of auditory profiles in auditory research, several limitations remain. First, the factors influencing the generation of auditory profiles remain largely unexplored. Elkhouly et al. (2021) investigated the impact of the number of profiles, suggesting that eight or ten profiles might be optimal when using a database containing only audiograms. In contrast, Saak et al. (2022) proposed that 13 profiles were optimal based on model-based clustering applied to an audiological database that included both audiograms and supra-threshold measures. In addition, Parthasarathy et al. (2020) applied a Gaussian Mixture Model (GMM) to a population data set and identified four optimal profiles based on Bayesian Information Criterion (BIC) values. However, the optimal number of profiles varies across studies, highlighting the need for further investigation. Please note that only data-driven approaches can optimize the number of profiles, whereas expert-based auditory profiles rely on expert judgment to determine this number. In addition to the number of profiles, we hypothesize that the choice of clustering algorithm in data-driven approaches may also influence the generation of auditory profiles, as different studies have employed various algorithms. For instance, Bisgaard et al. (2010) primarily used vector quantization, while Parthasarathy et al. (2020) utilized Gaussian mixture models.

Second, to the best of our knowledge, only few studies have systematically and quantitatively compared different auditory profiling frameworks (Dimitrov et al., 2025; Xu et al., 2025). Consequently, it is unclear which framework is most suitable, as the choice likely depends on the specific aim of the profiling and the characteristics of the available data. Therefore, one of the objectives of this paper is to address this gap by comparing different auditory profiles using both statistical and machine learning-based models based on the same patient database and to ultimately recommend the most suitable auditory profiling framework for a given purpose. A unique aspect of the present study is that all auditory profiling methods were applied and evaluated using the same large, openly accessible dataset, namely the extended Oldenburg Hearing Health Record (Jafri et al., 2025). This unified framework enables direct and fair comparisons across profiling strategies—a desirable feature pointed out in previous work (Dimitrov et al., 2025).

The comparison of different auditory profiling frameworks considers two key aspects: internal consistency, which reflects the extent to which individuals within the same profile are grouped together (i.e., low within-group variance), and external distinctiveness, which reflects how well individuals from different profiles are separated (i.e., high between-group variance). Overall, we aim at establishing an objective framework in fairly comparing different auditory profiling frameworks. Specifically, we propose employing statistical-based intrinsic measures (e.g., the Davies-Bouldin index; Davies & Bouldin, 1979), principal component analysis (PCA; Murphy, 2012), and manifold learning techniques such as t-distributed stochastic neighbor embedding (t-SNE; Van der Maaten & Hinton, 2008) to perform these comparisons. PCA and t-SNE are employed to reduce the high dimensionality of the original data set, facilitating visual comparison of different groups in a 2D space. This approach addresses the challenge of visualizing individual participants and comparing groups in the original high-dimensional data. Here, dimensionality reduction is used as an analytical tool to compare auditory profiling frameworks rather than as a reduced model of auditory functioning.

In summary, this study explores how methodological choices affect auditory profiling and how resulting profiles can be evaluated. We aim to address the following research questions, guided by the assumption that both the construction and evaluation of auditory profiles significantly influence their clinical and scientific utility:

1. How consistent and distinctive are the profiles produced by different auditory profiling frameworks?
2. How do the number of profiles and the clustering algorithm influence these outcomes?

3. How can different auditory profiling frameworks be compared using statistical or machine learning-based approaches to assess their effectiveness in capturing meaningful individual differences?

These questions build on the premise that clearer, data-driven criteria are needed to improve the reproducibility and comparability of auditory profiles across studies.

**Materials and methods**

*Overview of the data set*

The data set employed in this study was an extension of the Oldenburg Hearing Health Record (OHHR; Jafri et al., 2025). The extended data set consisted of 1127 participants with a mean age of 67.2 years (SD = 12.0), of whom 55.7% were male and 44.3% female. All participants underwent a comprehensive auditory evaluation comprising eight different tests. These included a questionnaire, two cognitive assessments (verbal intelligence and the DemTect test), the SF-12 health survey, an adaptive categorical loudness scaling test, two speech-in-noise tests (the Goettingen Sentence Test (GÖSA) and the Digit Triplet Test (DTT)), and pure-tone audiometry. Collectively, 37 parameters were derived from these tests. Air-conduction audiograms were assessed in sound-treated booths by hearing aid acousticians in accordance with IEC 60645-1 (2002). Measurements covered 11 audiometric frequencies ranging from 0.25 to 8 kHz for both ears, using a Unity II audiometer and HDA200 headphones. For further details on the measurements, see Gieseler et al. (2017), Saak et al. (2022, 2025), and Jafri et al. (2025). The present study demonstrates how a comprehensive data set which includes the open-access OHHR can be used to derive and compare data-driven auditory profiles.

*Applying auditory profiles to the data set*

Following the classification frameworks proposed by Bisgaard et al. (2010), Humes (2019), Cruickshanks et al. (2020), Dubno et al. (2013), Parthasarathy et al. (2020), Sanchez-Lopez et al. (2020), and Saak et al. (2022), participants were assigned to auditory profiles based on the parameters specified by each framework, which may include audiometric thresholds and, in some cases, additional supra-threshold measures. Table 2 summarizes the distribution of participants for the eight auditory profiling frameworks. Please note that 704, 545, and 87 participants were categorized as unidentified under the audiometric phenotype, general phenotype, and BEAR auditory profiles, respectively.

Table 2. Number of participants within each class of the respective auditory profiling framework. Please refer to Table 1 with respect to the details of the respective framework.

| Auditory profiling framework | Number of participants |
| --- | --- |
| Baseline | NH: 287, HI: 840 |
| Bisgaard profile | N1: 283, N2: 200, N3: 253, N4: 73, N5: 27, N6: 7, N7: 2, S1: 140, S2: 88, S3: 54 |
| WHO HI grades | Normal: 256, Mild: 312, Moderate: 327, Moderately severe: 160, Severe: 63, Profound: 9 |
| WARHICS levels | Level 1: 80, 2: 55, 3: 92, 4: 224, 5: 232, 6: 138, 7: 304, 8: 2 |
| Phenotype | O-N: 164, PRE-MET: 105, MET: 98, SENS: 20, MET + SENS: 36, Unidentified: 704 |
| General phenotype | Normal: 119, Flat: 36, HFHL: 240, Mixed: 187, Unidentified: 545 |
| BEAR | A: 431, B: 146, C: 395, D: 68, Unidentified: 87 |
| Hearing4all | Level 1: 40, 2: 18, 3: 36, 4: 21, 5: 66, 6: 257, 7: 67, 8: 20, 9: 25, 10: 291, 11:190, 12: 50, 13: 46 |

*Normalized intrinsic measures*

Following Elkhouly et al. (2021), three statistical intrinsic measures—Davies-Bouldin (DB) score, Calinski-Harabasz (CH) score, and Silhouette Index—were employed to evaluate the performance of the eight auditory profiling frameworks. All of the measures assess both internal consistency (i.e., similar individuals are grouped together) and group distinctiveness (i.e., distinct groups differ clearly from each other) (Davies & Bouldin, 1979; Caliński & Harabasz, 1974; Rousseeuw, 1987; Elkhouly et al., 2021). Lower Davies–Bouldin (DB) scores indicate better profiling performance, with an optimal value of 0, whereas higher Calinski–Harabasz (CH) and Silhouette Index values reflect superior auditory profiles; the Silhouette Index ranges from −1 to 1, with higher values indicating better performance.

To estimate the distribution of the three intrinsic measures, we performed bootstrapping by drawing 1,000 samples with replacement from the data set, each containing $N = 1{,}000$ data points. The calculations for all three intrinsic measures were carried out using the Python "scikit-learn" package and detailed descriptions of the intrinsic measures can be found in the supplementary material.

To enable a fair comparison across different auditory profiling frameworks, we normalized the intrinsic clustering measures with respect to the number of profiles.

Specifically, the DB score was normalized by dividing the original score by $\log_2(N)$, where N denotes the number of profiles. For the CH score and Silhouette Index, normalization was performed by multiplying each measure by $\log_2(N)$. The resulting values are referred to as the normalized DB score, normalized CH score and normalized Silhouette Index, respectively.

### *Factors influencing the generation of auditory profiles (clustering approach, number of profiles)*

We investigated two factors that could influence the clustering result of an auditory profiling framework if applied to the same data set, i.e., the clustering approach and the number of profiles. For the clustering approach, we compared two widely used methods: vector quantization (VQ), employed in the Bisgaard profiles (Bisgaard et al., 2010), and the Gaussian Mixture Model (GMM), applied in the General phenotype framework by Parthasarathy et al. (2020). The comparison was based on the Davies-Bouldin (DB) score. For comparability, both clustering methods were applied to the same dataset with the number of profiles fixed at 10. This choice was driven by methodological constraints: the number of Bisgaard profiles is predefined, whereas only the number of profiles in the General Phenotypes framework could be adjusted via the GMM. A t-test was performed to evaluate the impact of the clustering method on the DB score. To assess the effect of the number of profiles, we used the GMM method since its parameters could be freely adjusted to generate auditory profiles, and varied the number of profiles from 2 to 15. The resulting DB scores were plotted as a function of the number of profiles for comparison. The GMM model was implemented using the "scikit-learn" package in Python.

*Applying principal component analysis (PCA) on the extended Oldenburg Hearing Health Record (OHHR)*

Principal Component Analysis (PCA), an unsupervised machine learning method, was applied to reduce the dataset's dimensionality, visualize individual participants in a two-dimensional PCA space, and visually compare different auditory profiles (Murphy, 2012). Clustering methods (e.g., vector quantization, Gaussian mixture models) were used within these auditory profiling frameworks to group listeners based on shared auditory characteristics. The analysis was conducted using the R package "FactoMineR" (Lê et al., 2008), which employs singular value decomposition (SVD) to extract the first two principal components (PC1 and PC2). The "factoextra" package (Kassambara & Mundt, 2017) was used to generate PCA plots, with PC1 and PC2 displayed on the x- and y-axes, respectively. Before performing PCA, the data were standardized to have unit variance. The final analysis retained five dimensions ($N = 5$).

*Applying t-distributed stochastic neighbor embedding (t-SNE) on the extended Oldenburg Hearing Health Record (OHHR)*

t-Distributed Stochastic Neighbor Embedding (t-SNE) is an unsupervised manifold learning technique used to reduce data set dimensionality and visualize high-dimensional data, making it suitable for visually comparing different auditory profiles (Van der Maaten & Hinton, 2008; Van der Maaten, 2014). Unlike Principal Component Analysis (PCA), which preserves the global structure, t-SNE focuses on maintaining the neighborhood relationships of data points when mapping from high-dimensional to low-dimensional spaces. Additionally, t-SNE is a non-linear dimensionality reduction method, whereas PCA is linear.

    The t-SNE analysis was conducted using the R package "Rtsne" (Krijthe et al., 2018). The perplexity parameter was set to 30 (Van der Maaten, 2014), and the theta

parameter, controlling the trade-off between speed and accuracy, was set to 0.0. No initial PCA step was performed, and all other hyperparameters were kept at their default values (Krijthe et al., 2018). The two output variables from the t-SNE analysis were denoted as t-SNE Dimension 1 and t-SNE Dimension 2.

**Results**

*Impact of the clustering approaches and number of profiles*

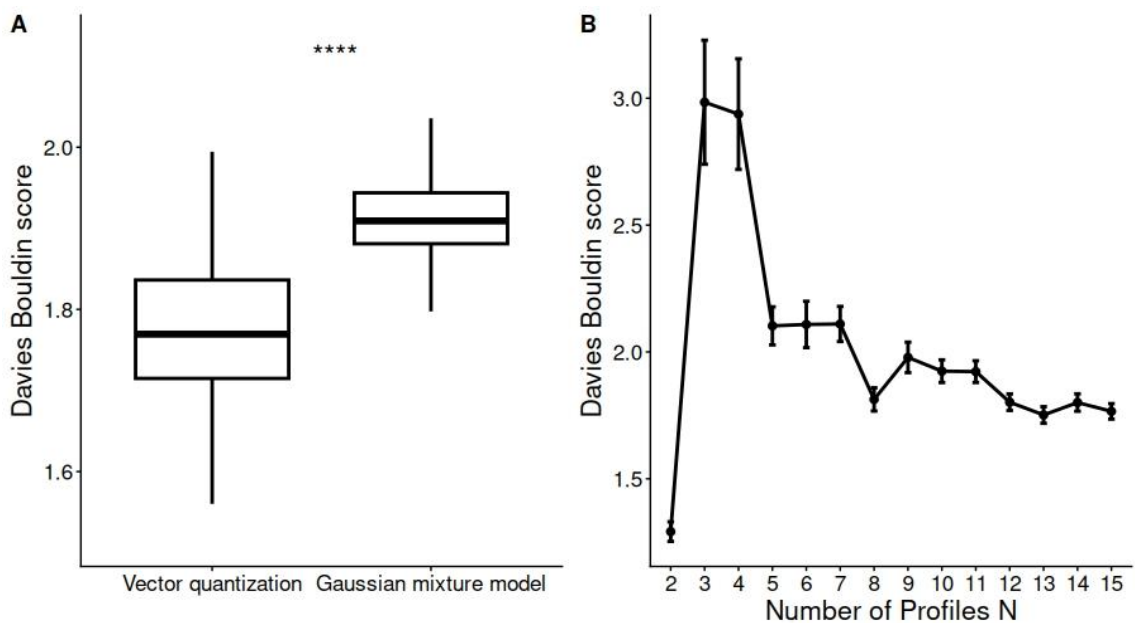

Fig. 1 A.) Box-plot of the Davies-Bouldin (DB) scores for two clustering approaches. B) the DB scores as a function of number of profiles N, varied using a generic Gaussian Mixture Model (GMM)-based clustering approach. Lower DB scores indicate better clustering performance, reflecting more compact and well-separated auditory profiles. Bar-and-whisker plots represent the median, 25th and 75th percentiles, and interquartile ranges (IQR). Whiskers extend to the most extreme values within $1.5 \times$ IQR from the 25th and 75th percentiles. Significance levels are denoted by stars (* for $p < 0.05$, ** for $p < 0.01$, *** for $p < 0.001$, **** for $p < 0.0001$).

Fig. 1A illustrates the effect of different clustering approaches on clustering performance. The DB scores differed significantly between the vector quantization method applied to Bisgaard profiles (Bisgaard et al., 2010) and the Gaussian Mixture

Model (GMM) used in the General phenotype (Parthasarathy et al., 2020), as confirmed by a t-test ($p < 0.05$). For this comparison, the number of profiles was fixed at 10. Fig. 1B shows the DB scores as a function of the number of profiles N when using the GMM for generating the profiles. Generally, as N increased, the DB scores initially increased and then gradually decreased. When N = 2, the DB score was at its lowest, indicating the best clustering performance. Excluding N = 2, the lowest DB score occurred at N = 13, which aligns with the findings by Saak et al. (2022). Conversely, when N = 3, the DB scores peaked, indicating the poorest clustering performance. As confirmed by ANOVA ($p < 0.05$), the DB scores shown in Fig. 1B differed significantly across N = 2–15.

*Normalized intrinsic measures*

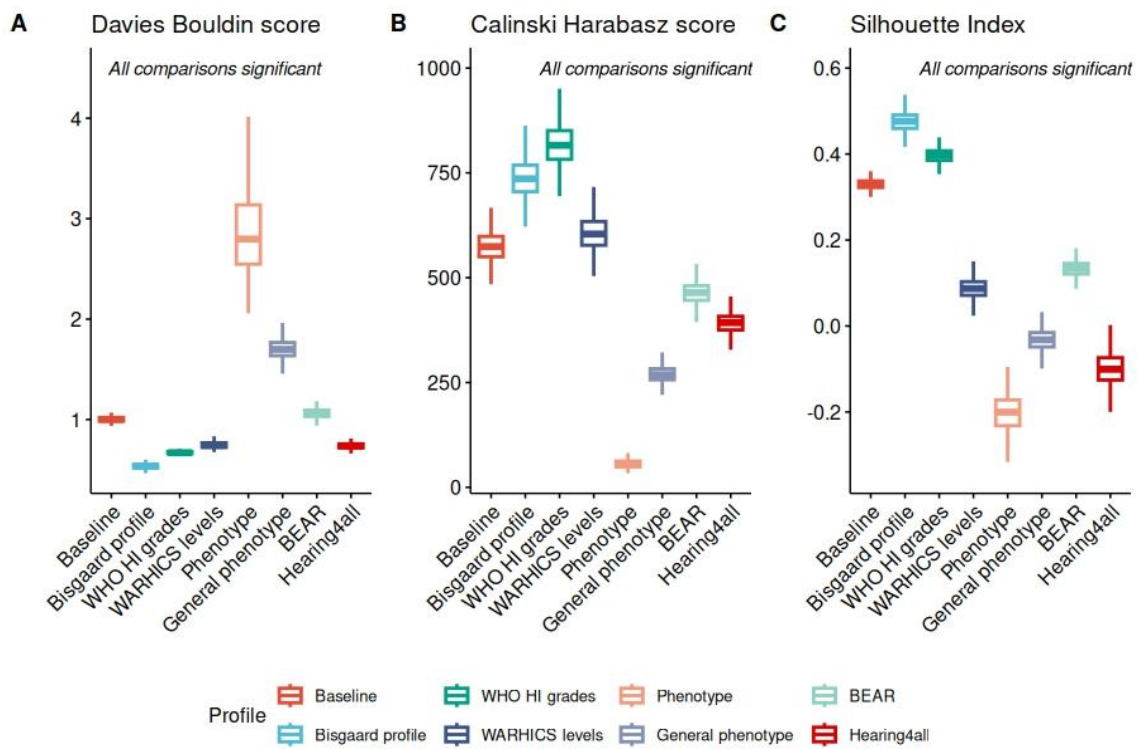

Fig. 2. Comparisons between eight auditory profiling frameworks (i.e., Baseline, Bisgaard profile, WHO HI grades, WARHICS levels, Phenotype, General phenotype, BEAR, and Hearing4all auditory profiles) in terms of three normalized intrinsic

measures (A. Davies-Bouldin (DB) score, B. Calinski-Harabasz (CH) score, and C. Silhouette Index). The smaller DB score, larger CH score, and Silhouette Index values closer to 1.0 indicate better performance in terms of internal consistency (i.e., similar individuals are grouped together) and well-separation (i.e., distinct groups differ clearly from each other). See Fig. 1 for an explanation of the box-plot.

Fig. 2A shows the normalized DB scores for different auditory profiling frameworks. Overall, the Bisgaard profile achieved the lowest normalized DB score, indicating the best performance, while the audiometric phenotype had the highest normalized DB score, reflecting the poorest performance. The normalized DB scores of the Bisgaard profile, WHO hearing impairment (HI) grades, WARHICS levels, BEAR, and Hearing4all auditory profiles were comparable, whereas the General phenotype scored slightly higher. In addition, the Hearing4all auditory profiles resulted in significantly lower normalized DB scores compared to the baseline auditory profile. Statistical analysis using ANOVA and post-hoc t-tests confirmed significant differences in normalized DB scores between all the auditory profiles ($p < 0.05$).

Figs. 2B and 2C present the normalized Calinski-Harabasz (CH) scores and Silhouette Index values, respectively. The highest normalized CH score was observed for the WHO HI grades, while the normalized Silhouette Index values for the Bisgaard profile were closest to 1.0. These results suggest that both the WHO HI grades and the Bisgaard profile effectively distinguish participants across groups and cluster similar participants together. This aligns well with the results shown in Fig. 2A, where both the WHO HI grades and the Bisgaard profile exhibited relatively low normalized DB scores. Conversely, the audiometric phenotype exhibited the lowest normalized CH score and the normalized Silhouette Index value farthest from 1.0, suggesting comparatively reduced clustering separation under the chosen profile configuration. Among the other frameworks, the baseline auditory profile, WARHICS levels, and BEAR profiles performed comparatively well, while the General phenotype performed slightly worse. In addition, the Hearing4all profiles exhibited lower CH and Silhouette Index values, suggesting comparatively reduced separation between profiles. ANOVA and post-hoc t-tests confirmed that the differences in all evaluation measures among the frameworks were statistically significant ($p < 0.05$).

*Principal component analysis (PCA)*

Fig. 3 compares the eight auditory profiling frameworks using principal component analysis (PCA). Nearly all auditory profiling frameworks could reliably identify normal-hearing participants on the left side of the subplots (e.g., the NH group in Figure 3A, the N1 group in Figure 3B, the Normal group in Figure 3C, and Group 1 in Figure 3D). Fig. 4 shows the contributing factors to the first and second principal components (PC1 and PC2). Overall, the PC1 and PC2 accounted for 38.2% and 9.7% of the variance in the data set, respectively. The factor loadings illustrated in Fig. 4A exhibit that the top six contributing factors to PC1 were all audiometric thresholds (e.g., pure-tone average and thresholds at various frequencies), indicating that PC1 primarily reflects hearing sensitivity at the threshold level. This coincides with the observation, that the normal hearing group is represented on the left and the group with the largest hearing loss on the right side of the respective subplot in Fig. 3. In contrast, Fig. 4B shows that the main contributors to PC2 were supra-threshold parameters, derived from the adaptive categorical loudness scaling test. Therefore, PC1 may represent deficits in audibility, whereas PC2 appears to capture aspects of supra-threshold auditory processing.

In Figs. 3A, 3C, and 3D, distinct group separations were observed for the baseline auditory profiles, WHO hearing impairment grades, and WARHICS levels. However, in Fig. 3B, Bisgaard profiles showed significant overlap, particularly between

N-type and S-type audiograms (e.g., N1 and S1), suggesting that PCA may not effectively capture certain features, such as audiogram slope.

Figs. 3E and 3F depict PCA results for audiometric phenotypes and general phenotypes. In Fig. 3E, a large number of participants remained unclassified, and all groups appeared heavily mixed. This aligned with the quantitative results in Fig. 2, where the audiometric phenotype framework performed the worst. Similarly, in Fig. 3F, many participants could not be assigned to any general phenotype, although clear group separations were observed among the remaining participants.

In Fig. 3G, only a few participants could not be categorized into one of the four BEAR auditory profiles. However, our results differed from those reported by Sanchez-Lopez et al. (2018; 2020). Their study demonstrated clear group separations, with profiles A, B, C, and D occupying distinct quadrants of the PCA plot, while unidentified participants clustered centrally (see Fig. 3 in Sanchez-Lopez et al. 2020 for details). In contrast, our analysis showed overlapping groups. This discrepancy may stem from differences in the data sets and the examination procedures between the two studies. Our cohort was larger and potentially more diverse, while Sanchez-Lopez et al. (2020) may have focused on more targeted archetypes. In addition, Sanchez-Lopez et al. included specific tests like the fast spectro-temporal modulation test (Bernstein et al., 2016) and binaural pitch processing (Santurette & Dau, 2012), which were not part of our test battery. However, these specific tests played a limited role in the classification of auditory profiles in Sanchez-Lopez et al. (2020), indicating that their omission can not explain the discrepancy observed here. Figure 3H illustrates the 13 Hearing4all auditory profiles. These profiles also exhibited considerable overlap and mixing.

In summary, while the baseline auditory profile, WHO hearing impairment grades, and WARHICS levels provided relatively distinct groupings, the audiometric phenotype framework appeared less consistent. However, it is important to emphasize that the usefulness of a classification framework should be judged not only by the clarity of separation but by the clinical or interpretative value of the resulting profiles. For example, although substantial participant overlap is observed within the audiometric phenotype framework—suggesting comparatively limited discriminative power relative to other profiling approaches—it nevertheless may retain important clinical value. Specifically, audiometric phenotypes are motivated by hypothesized mechanisms of hearing loss and serve as descriptive frameworks for organizing audiometric patterns. One limitation of the PCA analysis is that restricting the representation to two dimensions may be insufficient to clearly distinguish among different profiles; future analyses could therefore incorporate additional dimensions.

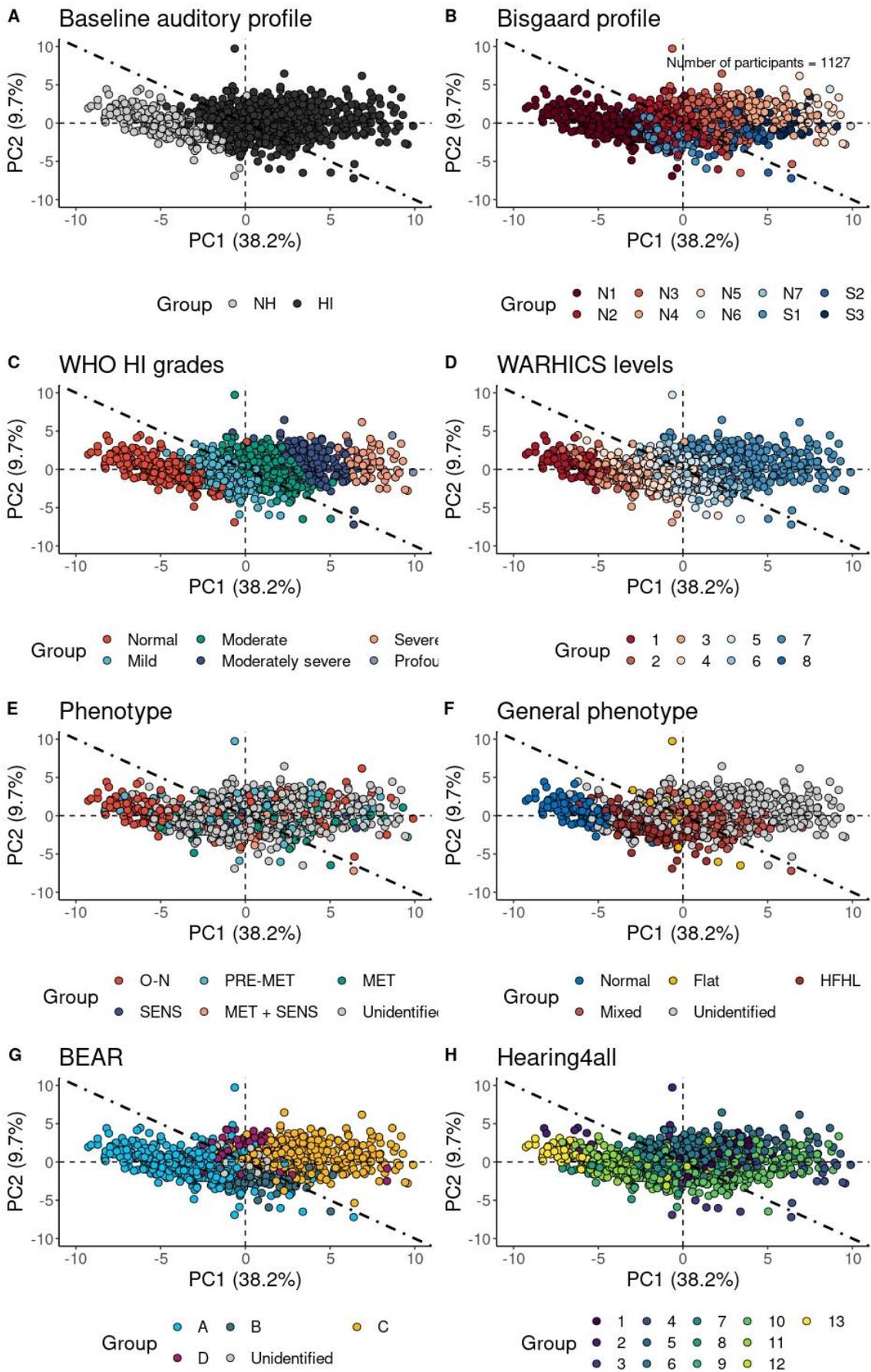

Fig. 3. Principal Component Analysis (PCA) plots for eight different auditory profiling frameworks based on data from N = 1127 participants. The x-axis represents the first principal component (PC1), and the y-axis represents the second principal component (PC2). The same PCA is applied across all panels. In each panel, colors represent group assignments according to a specific auditory profiling framework. The dot-dashed line indicates the diagonal. A.) Baseline auditory profile (NH: Normal hearing; HI: Hearing impaired). B.) Bisgaard profile (Bisgaard et al., 2010). C.) WHO hearing impairment (HI) grades (Humes, 2019). D.) WARHICS levels (Cruickshanks et al., 2020). E.) Audiometric phenotypes (Dubno et al., 2013). F.) General phenotype classification (Parthasarathy et al., 2020). G.) BEAR auditory profiles (Sanchez-Lopez et al., 2020). H.) Hearing4All auditory profiles (Saak et al., 2022).

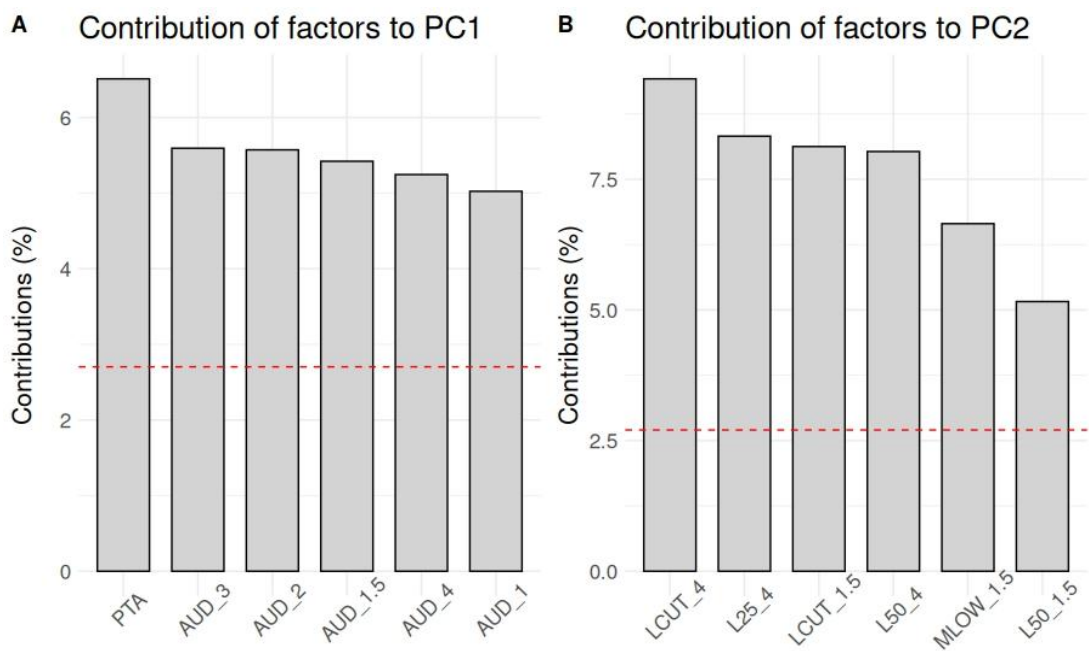

Fig. 4. Contributions of factors to (A) the first principal component (PC1) and (B) the second principal component (PC2). Only the top six contributing factors are shown. Red dashed line: average contribution. Sub-figure A: PTA – pure-tone average; AUD_3/2/1.5/4/1 – thresholds at 3, 2, 1.5, 4, and 1 kHz, respectively. Sub-figure B: LCUT_4/1.5 – transition levels of loudness growth functions at 4 and 1.5 kHz; L25_4 – medium loudness level (MLL) at 4 kHz; L50_4/1.5 – uncomfortable loudness level (UCL) at 4 and 1.5 kHz; MLOW_1.5 – slope at low levels at 1.5 kHz (see Oetting et al.

(2014) for the explanation of these supra-threshold parameters derived from the loudness growth functions).

*t-distributed stochastic neighbor embedding (t-SNE)*

Figs. 5A–5H present the results of t-Distributed Stochastic Neighbor Embedding (t-SNE), a non-linear manifold learning method, applied to eight auditory profiling frameworks. The x-axis and y-axis represent t-SNE dimensions 1 and 2, respectively. Unfortunately, unlike PCA, t-SNE does not produce axes with clear, interpretable meaning (Faust et al., 2017). Nevertheless, a similar rough annotation of the dimension 1 with overall hearing loss (in inverted order) is discernible as with the PCA (cf., Fig. 3) since the normal hearing class is on the extreme right in all subplots and the class with the highest average hearing loss on the left. A more detailed comparisons among these auditory profiling frameworks shows strong alignment with the PCA results as well (see Fig. 3). Specifically, groups in the baseline auditory profile, WHO HI grades, and WARHICS levels were clearly separable, whereas groups in the audiometric phenotypes were more intermixed. Comparing Figs. 3 and 5, t-SNE appeared to offer slightly better separation of groups. For example, in the BEAR auditory profile, Groups B and D were distinct from Groups A and C in Fig. 5G, while in Fig. 3G, Groups B and D were mixed with Groups A and C. These results suggest that t-SNE outperforms PCA in separating groups in low-dimensional space, as expected. Unlike PCA, which projects data based on linear variance, t-SNE preserves the relationships between data points, particularly the local neighborhood distances, in the low-dimensional space. Consequently, t-SNE may be a more suitable method than PCA for dimensionality reduction in this context. However, limiting the analysis to two dimensions may obscure class separability that exists in higher-dimensional representations, indicating that future work should examine additional dimensions when applying PCA and t-

SNE.

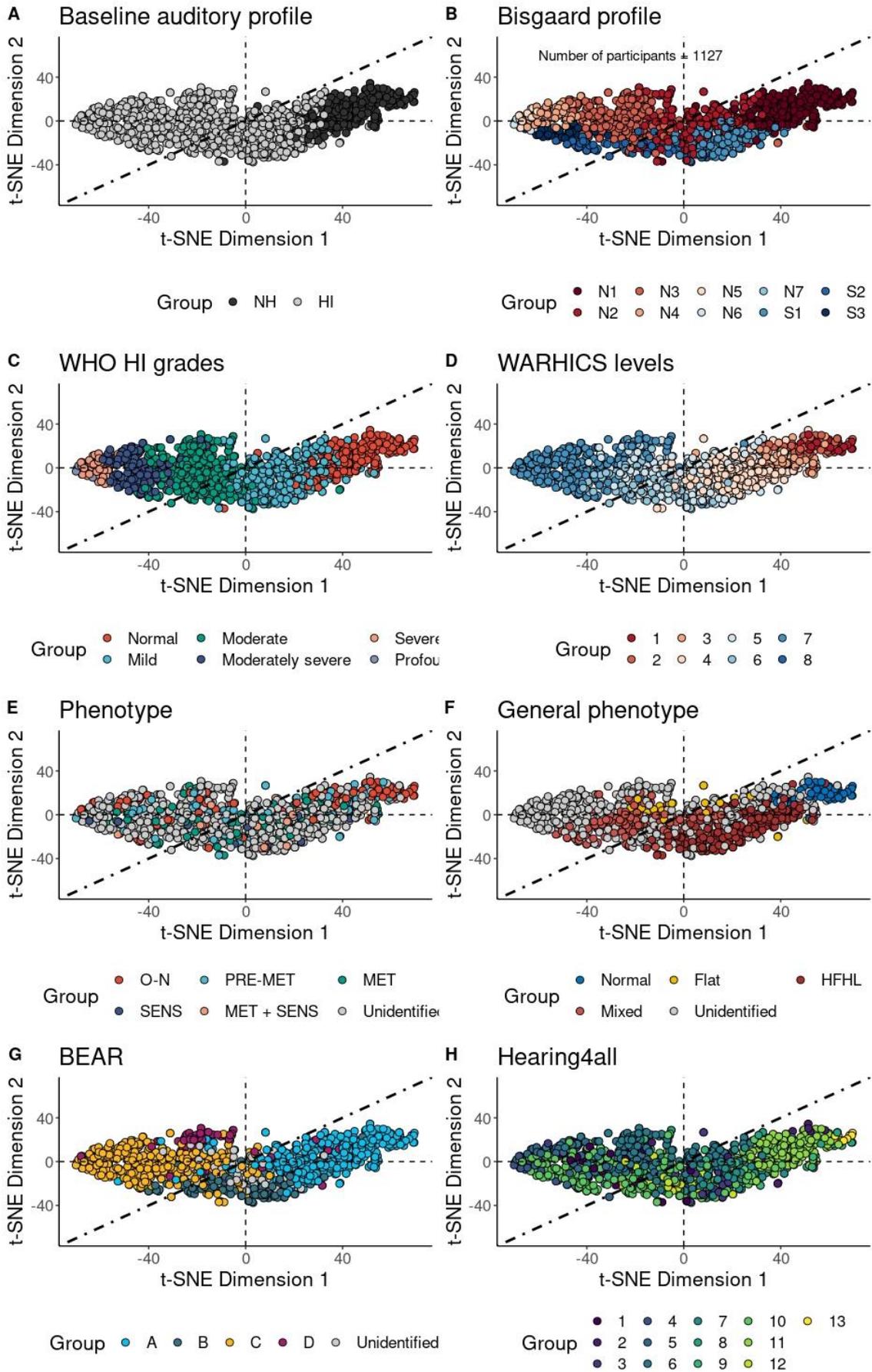

Fig. 5. t-Distributed Stochastic Neighbor Embedding (t-SNE) results for N=1127 participants. A.) Baseline auditory profile (NH: Normal hearing; HI: Hearing impaired). B.) Bisgaard profile (Bisgaard et al., 2010). C.) WHO hearing impairment (HI) grades (Humes, 2019). D.) WARHICS levels (Cruickshanks et al., 2020). E.) Audiometric phenotypes (Dubno et al., 2013). F.) General phenotype classification (Parthasarathy et al., 2020). G.) BEAR auditory profiles (Sanchez-Lopez et al., 2020). H.) Hearing4All auditory profiles (Saak et al., 2022). The x-axis represents t-SNE dimension 1, and the y-axis represents t-SNE dimension 2. Dot-dashed line: the anti-diagonal. Group variables depicted in the sub-figures correspond to those in Fig. 3.

**Discussion**

*Factors influencing the generation of auditory profiles*

We systematically investigate two major factors that influence the generation of auditory profiles, using an extension of the Oldenburg Hearing Health Record (OHHR) as a common data set, enabling direct and fair comparisons across approaches — an advantage not achievable with literature-based values alone.

First, the clustering approach plays a significant role, aligning with our expectations. Using the same number of profiles (N = 10) and the same data set, the vector quantization (VQ) method employed by Bisgaard et al. (2010) outperforms the Gaussian mixture model (GMM) used by Parthasarathy et al. (2020) with respect to separability between auditory profiles, as evidenced by significantly lower DB scores (see Fig. 1A). The primary distinction between VQ and GMM lies in their handling of cluster overlap: VQ enforces hard cluster boundaries, while GMM allows for overlapping clusters (Kinnunen et al., 2011). This difference may partly account for the observed advantage of VQ in our study, although further validation is needed to confirm this finding. Furthermore, VQ is a simpler, non-parametric approach, whereas GMM is a parametric method with many free parameters, making it computationally intensive.

GMM also assumes that the data follow a Gaussian mixture distribution, an assumption that may not hold in our case. In future studies, it could be insightful to compare these approaches with the decision tree method used by Sanchez-Lopez et al. (2020), as well as to evaluate their performance across independent datasets.

Second, the number of profiles significantly affects the generation of auditory profiles. Our findings align with those of Saak et al. (2022), indicating that a profile count of 13 yields optimal clustering performance (see Fig. 1B). However, we disagree with Elkhouly et al. (2021) and Parthasarathy et al. (2020), who suggested optimal profile counts of 10 and 6, respectively. These discrepancies may arise from differences in evaluation metrics: we used DB scores, Elkhouly et al. (2021) employed Silhouette Index values, and Parthasarathy et al. (2020) applied the Bayesian Information Criterion (BIC). Additionally, the clustering methods differ between studies: we and Parthasarathy et al. (2020) used GMM, whereas Elkhouly et al. (2021) employed spectral clustering (SP). Also the datasets underlying the profiles were different. Future research should investigate additional factors, such as data set composition and participant numbers, that could influence the generation of auditory profiles.

### *Comparison between auditory profiles*

The intrinsic clustering scores used in this study quantify complementary aspects of profile structure, such as compactness and separation, and are intended to inform—rather than determine—the design of auditory profiling frameworks. Profiles with clearer separation and internal consistency may support more precise diagnostics by reducing ambiguity in profile assignment and enabling more differentiated treatment strategies. However, higher intrinsic scores alone do not guarantee clinical usefulness, as interpretability and actionable meaning remain essential. Consequently, these scores should be interpreted with care as guiding trade-offs between differentiation and clinical applicability in the context of precision audiology.

In this study, we quantitatively compared eight auditory profiling frameworks using intrinsic measures and visually analyzed them through principal component analysis (PCA) and t-distributed stochastic neighbor embedding (t-SNE). These normalized intrinsic measures assess the auditory profiling frameworks in two key aspects: the extent to which participants with the same profile cluster together, and how well participants with different profiles are separated. This is an important consideration, as an effective auditory profiling framework should aim to maximize between-group variance while minimizing within-group variance. Our findings demonstrate that these approaches can effectively compare auditory profiling frameworks and identify the optimal auditory profile. In addition, the same data set—the extended Oldenburg Hearing Health Record (OHHR) data set—is used in this study to perform auditory profiling using different approaches and to compute various intrinsic measures for the resulting profiling frameworks.

As anticipated, the baseline auditory profile tended to perform well in separating participants into two groups based on their pure-tone averages (PTA4), a strategy commonly used in the field, likely because it reflects a binary classification scheme, which makes group separation between a very homogeneous normal group and an inhomogeneous group with hearing impairment rather trivial. This finding is supported by Fig. 1B, which shows that when $N = 2$, the Davies-Bouldin (DB) scores are at their lowest, indicating the highest clustering performance. However, despite this advantage, baseline auditory profiles provide limited support for achieving more precise diagnostics and optimized treatment within a precision-audiology framework. Hence, further exploration of subgroups within hearing-impaired (HI) listeners is necessary, as they are not homogeneous, particularly regarding supra-threshold measurements such as the loudness scaling data reported in Xu et al. (2024b), which are not included in most of the best-performing threshold-based profiling frameworks.

Across all auditory profiling frameworks, the audiometric phenotype shows the weakest ability to differentiate between groups of listeners. This may be attributed to both within-profile and between-profile variations. On the one hand, the within-profile variations are substantial, as evidenced by participants sharing the same audiometric phenotype (e.g., the older-normal phenotype) who are dispersed across the PCA and t-SNE 2-D space (see Figs. 3E and 5E). On the other hand, the between-profile differences are minimal. For instance, schematic boundaries between phenotypes overlap (as illustrated in Fig. 1 of Dubno et al. (2013)), and shared characteristics are evident across phenotypes (as shown in Table 1 of Dubno et al. (2013)).

For the Bisgaard profiles, all approaches consistently indicate that participants with N-type or S-type audiograms are relatively well-clustered. However, distinguishing between N-type and S-type audiograms remains challenging, as shown in Figs. 3B and 5B. In Bisgaard et al. (2010), the PTA4 value for N1 is 10 dB HL, which is very close to the PTA4 for S1 (12 dB HL). Similarly, N2's PTA4 is comparable to S2's, and N4's is similar to S3's. These comparable PTA4 values, despite the differing slopes of the audiograms, result in small between-profile differences, which likely contribute to the difficulty in differentiating between N-type and S-type audiograms in the reduced two-dimensional space.

Both WHO Hearing Impairment (HI) grades and WARHICS levels demonstrate strong performance in categorizing participants into distinct auditory profiles. This is evidenced by their low normalized Davies-Bouldin (DB) scores, normalized high Calinski-Harabasz (CH) scores (Fig. 2), and the clustering patterns observed in the PCA and t-SNE plots (Figs. 3C–3D and 5C–5D). Please note that some groupings appear comparable across the Bisgaard profiles, WHO HI grades, and WARHICS levels. For instance, the N3 Bisgaard profile aligns with the "moderate" WHO HI grade and WARHICS level 6.

These findings suggest a degree of correspondence between different auditory profiling frameworks, which is also reflected in Table 1—for example, WHO HI and WARHICS are both expert-based approaches grounded in epidemiological audiogram data—partly explaining their relatively strong performance in the two-dimensional space, where one axis largely represents audiometric information. For example, Humes et al. (2021) compared WARHICS levels with WHO HI grades and observed that several WARHICS levels fell within specific WHO HI grades. They argued that WARHICS levels offer a finer categorization than WHO HI grades. However, our results suggest a more nuanced picture. The lower normalized CH scores and normalized Silhouette Index observed for WARHICS levels indicate less distinct or compact clustering in our data, potentially reflecting overlapping group boundaries or reduced separability. This may imply that, despite offering more categories, the WARHICS framework does not necessarily enhance cluster quality in our sample.

The relationship between WHO HI grades and Bisgaard profiles can also be inferred through PTA4 (pure-tone average across 500 Hz, 1 kHz, 2 kHz, and 4 kHz). Bisgaard et al. (2010) established the link between PTA4 and Bisgaard profiles in Table 1 of their study, while Humes et al. (2019) reported the corresponding relationship between PTA4 and WHO HI grades in Table 1 of their work. This allows for a comparative analysis of Bisgaard profiles and WHO HI grades via PTA4, further illustrating the interconnectedness of these frameworks.

Compared to the audiometric phenotype, the general phenotype demonstrates better clustering of participants into distinct groups, as indicated by a smaller normalized DB score, a larger normalized CH score, and a normalized higher SI (see Fig. 2 for details). As illustrated in Fig. 3C of Parthasarathy et al. (2020), the audiogram boundaries for the general phenotype are broader—particularly at frequencies below 2 kHz—compared to the audiometric phenotype (see Fig. 1 in Dubno et al., 2013). This broader range suggests that the general phenotype achieves slightly better clustering performance due to its less restrictive classification criteria. Additionally, there is a clear correspondence between the two phenotypes: Older-normal, Metabolic, Sensory, and Metabolic + Sensory audiometric phenotypes align with the Normal, Flat, High Frequency Hearing Loss (HFHL), and Mixed general phenotypes, respectively.

The relatively low normalized DB score, high normalized CH score, and high normalized Silhouette Index suggest that the BEAR auditory profiles effectively segregate participants. Sanchez-Lopez et al. (2020) compared the four BEAR auditory profiles with the audiometric phenotypes and found that Profiles A and B corresponded to the sensory audiometric phenotype, Profile D aligned with the metabolic audiometric phenotype, and Profile C represented a mixed phenotype (metabolic + sensory). However, this clear correspondence is not evident in Figs. 3 and 5, possibly because only a small number of participants were assigned to specific audiometric phenotypes, while the majority remained unidentified.

In contrast, the 13 Hearing4all auditory profiles significantly outperform the audiometric phenotypes but generally underperform compared to the BEAR profiles for two of the three intrinsic measures shown in Fig. 2. One confounding factor, however, is the comparatively large number of "unidentified" cases for the BEAR profiles (as opposed to no unidentified case for the Hearing4all profiles), which is expected to lead to a better clustering performance since the "difficult to classify" cases are left out. The Hearing4all profiles exhibit large within-profile differences, as indicated by the considerable interquartile ranges for certain features (e.g., the speech recognition threshold in Profile 6; see Figure 3 in Saak et al., 2022, for details). These discrepancies may contribute to the slightly weaker performance of the Hearing4all profiles relative to the BEAR profiles if a direct comparison is valid at all. However, the Hearing4all framework comprises a larger number of profiles, which may enhance its applicability for more fine-grained diagnostics and treatment planning and leads to a better Davies Bouldin score.

We concur with Saak et al. (2022) on many advantages of the Hearing4all auditory profiles, such as their flexibility in profile numbers, ease of extension, integration of Bisgaard profiles into the profiling process, and comprehensiveness, which ensures that all participants can be assigned to a derived profile. However, some limitations must be acknowledged, including the lack of expert validation and low interpretability. Although the Bisgaard profiles are explicitly incorporated as input for generating the Hearing4all profiles, the correspondence between the Bisgaard profiles—and other auditory profiling frameworks—and the Hearing4all profiles remains unclear.

### *Audiogram-based vs. comprehensive auditory profiles*

Given their higher number of input parameters characterizing the individual patients in a more comprehensive way, it is unexpected that the two comprehensive auditory profiles (i.e., the BEAR and Hearing4all auditory profiles) do not outperform audiogram-based auditory profiles in normalized intrinsic measures, PCA, and t-SNE. One possible reason is that the evaluation metrics used may not accommodate high-dimensional data representations and tend to favor simpler models. Another possible reason is that threshold and supra-threshold parameters do not necessarily vary together in a consistent way across auditory profiles with the consequence that the comprehensive auditory profiles may fail to consistently differentiate between participants. Even though threshold and supra-threshold measures are known to be at least partially correlated with each other, it is also well known from the literature that they carry information which can be independent of each other: For instance, participants with identical audiograms can exhibit markedly different supra-threshold parameters, such as speech recognition threshold (SRT; Balan et al., 2023) or loudness growth (Oetting et al., 2014; Xu et al., 2024b). Conversely, participants with different audiograms may share similar supra-threshold parameters, as seen with SRT (Hoppe et al., 2020).

Compared with comprehensive auditory profiles, audiogram-based profiles are simpler and easier to generate, as audiograms are routinely available in most clinical and research settings and have fewer independent parameters. In contrast, comprehensive auditory profiling requires additional supra-threshold listening measures, which are not yet standard in many clinics and may vary across laboratories. However, one of the central motivations for auditory profiling is a more detailed characterization of hearing loss that extends beyond threshold sensitivity. Accordingly, comprehensive profiles incorporate additional dimensions of supra-threshold processing, providing complementary information that cannot be captured by audiograms alone and is therefore essential for more precise characterization and individualized intervention. For instance, speech recognition in noise is supposed to be an important auditory function assessed both by the Hearing4all auditory profile (using the German Göttingen Sentence Test) and the BEAR auditory profile (using the Danish Hearing-in-Noise Test (HINT)). An integration into a common set of comprehensive auditory profiles (e.g., Saak et al., 2025) may provide a language-independent way of exploiting the results of these language-specific tests into the same framework even though the tests by themselves differ or may not be applicable to the same subject. Such an integration, however, requires a standardized test battery that includes both audiograms and supra-threshold tests to comprehensively assess auditory function (Van Esch et al., 2013). Additionally, a standardized data structure is essential to facilitate data sharing across institutions (Kollmeier et al., 2024) which can lead to a big-data-supported set of auditory profiles by exploiting diverse large datasets via federated learning (Saak et al., 2025). .

*Epidemiological vs. audiological auditory profiles*

The epidemiological auditory profile primarily includes the WHO hearing impairment (HI) grades and the WARHICS levels. These profiles are developed using population datasets with normative participants, while audiological auditory profiles are derived from clinical or audiological datasets (e.g., the Oldenburg Hearing Health Record; see Jafri et al., 2025). Epidemiological profiles generally encompass a broader range of participants in terms of age (including both young adults and older individuals) and have larger sample sizes, typically exceeding 10,000 participants. In contrast, audiological profiles focus on more targeted samples, such as individuals with hearing disorders or those aged 50–89 years, and are limited to smaller sample sizes, usually around 1,000 participants, which are substantial for research datasets but remain modest compared with large-scale clinical data.

Both epidemiological auditory profiles effectively differentiate participants within the current audiological data set, as indicated by a low normalized DB score, and high normalized CH score and normalized Silhouette Index. These findings are consistent with those of Humes et al. (2021) and demonstrate that these profiles are not only applicable to population datasets but can also be extended to clinical datasets. In contrast, the clustering performance of auditory profiles developed using audiological datasets varies significantly. For example, some profiles, such as the Bisgaard profile, perform well, while others, like the audiometric phenotype, perform poorly. These differences may stem from the audiological data set employed in the current study, i.e., an extension of the OHHR. Profiles developed using different audiological data sets with a different case mix which does not align well with the current data set, might therefore exhibit a suboptimal performance.

*Intrinsic measures, PCA, and t-SNE*

To date, our study is among the first to provide quantitative comparisons of different auditory profiling frameworks using intrinsic measures. Specifically, we employed three intrinsic metrics: the Davies-Bouldin (DB) score, the Calinski-Harabasz (CH) score, and the Silhouette Index, which generally produced consistent results.

The DB score evaluates clustering quality by calculating the ratio of within-profile distance to between-profile distance (see Formula S1). A smaller DB score indicates better clustering performance, reflecting minimized within-profile distances and maximized between-profile distances. Conversely, the CH score assesses clustering by computing the ratio of between-profile variance to within-profile variance (see Formula S2). Higher CH scores signify better clustering, indicating maximized between-profile variance and minimized within-profile variance. While both the DB and CH scores focus on evaluating clustering performance at the profile level, the Silhouette Index offers a more comprehensive view. It provides insights at both the individual data point level and the overall profile level by averaging (see Formula S3). The Silhouette Index measures how similar a data point is to its own cluster compared to other clusters. Given that these three intrinsic measures capture different aspects of auditory profiling frameworks, we recommend their adoption in future research to provide a more robust evaluation of clustering frameworks than single-metric approaches. If different metrics do not agree, one should consider a combination of all metrics.

We compared different auditory profiling frameworks visually using Principal Component Analysis (PCA) and t-distributed Stochastic Neighbor Embedding (t-SNE), two dimensionality reduction techniques. Both methods reduce the original data set to two dimensions, enabling the visualization of participants across different groups in a derived 2D space for effective comparison. PCA, widely used in earlier studies for auditory profile comparisons (e.g., Sanchez-Lopez et al., 2018; 2020), provides a straightforward and computationally efficient approach. However, it is notably sensitive to outliers. In contrast, t-SNE, which is less commonly applied in this field, preserves the local structure of the data by maintaining the relative distances between points and is less affected by outliers. Our findings suggest that t-SNE offers better separation of participants compared to PCA. While PCA remains valuable for its simplicity and speed, t-SNE provides a complementary perspective, particularly in visualizing complex or locally structured data. Taken together, t-SNE should be considered an effective tool for visually comparing auditory profiles.

### *Limitations and outlook*

Currently, comparisons among different auditory profiling frameworks are based on the extension of a clinical database with open access, namely the Oldenburg Hearing Health Record. However, this data set primarily includes older patients with hearing impairments. While profiles derived from this data set offer valuable insights, their generalizability may be constrained. To ensure robust and unbiased findings, it is essential to validate these profiles using additional datasets, including population-based datasets. One suitable option is the National Health and Nutrition Examination Survey (NHANES), which has been extensively used in other studies (Ellis et al., 2021; Parthasarathy et al., 2020; Humes et al., 2023a, 2023b, 2023c). The NHANES data set includes over 10,000 participants, offering a broader spectrum of hearing abilities. Its test battery encompasses demographic questionnaires and audiometric evaluations but lacks many supra-threshold auditory assessments, such as speech-in-noise testing.

Second, the correspondence among different auditory profiling frameworks remains unclear. While some studies address this issue—such as Sanchez-Lopez et al. (2020), who explored the relationship between audiometric phenotypes and the BEAR auditory profile, and Saak et al. (2022), who examined links between the Bisgaard profile and the Hearing4all profile—a unified theoretical framework for correspondence would be invaluable. Such a framework could facilitate better understanding of the relationships among auditory profiling frameworks and enable their mutual conversion or estimation.

Third, our study demonstrates that two factors—the number of profiles and the clustering approach—significantly influence the generation of auditory profiles. Future research could investigate additional factors, such as the number of participants and potential interactions. An increased sample size may lead to the identification of more profiles, thereby affecting profile generation. Moreover, incorporating a broader range of auditory test features might further impact the development of auditory profiles.

Fourth, an objective validation tool for auditory profiles, particularly data-driven profiles, is lacking. Even though the current work provides some bench-marking with respect to three intrinsic measures as a first step towards validating the auditory profiling frameworks considered, it remains unclear which data-driven profiles are redundant or irrelevant and should therefore be excluded. Future research should focus on methods to merge similar profiles (Saak et al., 2025) and identify the most relevant and interpretable auditory profiles to ensure a streamlined and clinically meaningful framework.

**Conclusions**

Both factors under investigation significantly influenced the generation of auditory profiles: the clustering method and the number of profiles. Specifically, vector quantization outperformed the Gaussian mixture model by producing lower DB scores for this dataset with 10 profiles. Furthermore, as the number of profiles (N) increased from 2 to 15, clustering performance using the Gaussian mixture model initially decreased and then gradually improved. The best performance was observed with N ≈ 13, while N = 3 yielded the poorest results.

Eight auditory profiling frameworks were compared, including the baseline auditory profile, the Bisgaard profile, WHO hearing impairment (HI) grades, WARHICS levels, audiometric phenotype, general phenotype, the BEAR auditory profile, and the Hearing4all auditory profile. From these, the Bisgaard auditory profiles demonstrated the best performance in distinguishing participants into distinct groups, while the audiometric phenotype performed the worst. However, these relative differences should be interpreted in light of the specific analytical setting and may vary depending on the intended application and context. The clustering performance of the remaining six frameworks was comparable. From the two comprehensive auditory profiles under consideration, the BEAR auditory profile performed better than the Hearing4All profile for 2 out of 3 intrinsic measures, but at the cost of a significant percentage of data sets which had to be excluded and the disadvantage of providing only 4 different classes which limits its practical applicability. The Hearing4All auditory profile achieved with its optimum number of classes (N = 13) the lowest Davies Bouldin score among the comprehensive auditory profiles and should therefore be considered for future work.

Both PCA and t-SNE visualizations displayed participants in a two-dimensional space, enabling intuitive group-wise comparisons. Notably, t-SNE slightly outperformed PCA in differentiating participant clusters, as expected.

In summary, this study demonstrates that auditory profiling frameworks can be effectively and efficiently compared using manifold learning techniques and intrinsic measures. It presents the first quantitative comparison of these frameworks with the same underlying data, offering valuable insights into auditory profile classification and clustering methodologies.

**Acknowledgments**

This work was funded by the Deutsche Forschungsgemeinschaft (DFG, German Research Foundation) under Germany's Excellence Strategy – EXC 2177/1 - Project ID 390895286.

**Disclosure statement**

No potential conflict of interest was reported by the author(s).

**Data availability**

The data that support the findings of this study are openly available in Zenodo at http://doi.org/10.5281/zenodo.14177903.